\documentclass[preprint,nofootinbib,amsmath,amssymb,12pt]{revtex4}

\usepackage{cancel}
\usepackage{color}
\usepackage{slashed}
\usepackage{graphicx}
\usepackage{dcolumn}
\usepackage{bm}
\usepackage{hyperref}

\begin{document}

\title{Implications on the first observation of charm CPV at LHCb}

\author{Hsiang-nan Li$^1$\footnote{hnli@phys.sinica.edu.tw}, Cai-Dian L\"u$^2$\footnote{lucd@ihep.ac.cn}, Fu-Sheng Yu$^3$\footnote{yufsh@lzu.edu.cn}}

\address{
$^1$Institute of Physics, Academia Sinica, Taipei, Taiwan 11529, Republic of China\\
$^2$Institute of High Energy Physics, Chinese Academy of Sciences, Beijing 100049, People's Republic of China\\
$^3$School of Nuclear Science and Technology,  
Lanzhou University, Lanzhou 730000,  People's Republic of China}

\begin{abstract}
Very recently, the LHCb Collaboration observed the $CP$ violation (CPV) in the charm sector for the first time, with $\Delta A_{CP}^{\rm dir}\equiv A_{CP}(D^0\to K^+K^-)-A_{CP}(D^0\to \pi^+\pi^-)=(-1.54\pm0.29)\times10^{-3}$. This result is consistent with our prediction of $\Delta A_{CP}^{\rm SM}=(-0.57\sim -1.87)\times 10^{-3}$ obtained in the factorization-assisted topological-amplitude (FAT) approach in [PRD86,036012(2012)]. It implies that the current understanding of the penguin dynamics in charm decays in the Standard Model is reasonable. Motivated by the success of the FAT approach, we further suggest to measure the $D^+\to K^+K^-\pi^+$ decay, which is the next potential mode to reveal the CPV of the same order as $10^{-3}$.

\end{abstract}

\maketitle

\section{Introduction}
Very recently, the LHCb Collaboration observed the $CP$ violation (CPV) in the charm sector for the first time \cite{1903.08726}, with 
\begin{align}\label{eq:AcpLHCb2019}
\Delta A_{CP}\equiv A_{CP}(D^0\to K^+K^-)-A_{CP}(D^0\to \pi^+\pi^-)=(-1.54\pm0.29)\times10^{-3}.
\end{align}
This is a milestone of high energy physics, since CPV has been well established in the kaon and $B$ systems for many years, while the search for CPV in the charm sector has bot been successful until now. The difficulty of searching for charm CPV is due to its smallness in the Standard Model (SM). The naive expectation of CPV in charm decays is 
\begin{align}\label{eq:naiveAcp}
{\alpha_s(\mu_c)\over \pi}{|V_{ub}V_{cb}^*|\over |V_{us}V_{cs}^*|}=\mathcal{O}(10^{-4}).
\end{align}
Thus the charm CPV was usually a null test of the SM and a signal of new physics (NP) if it was observed to bef larger than percent level. In the late of 2011, the LHCb Collaboration reported an evidence of charm CPV at the order of percent \cite{LHCb2011},
$\Delta A_{CP}=(-8.2\pm2.4)\times10^{-3}$. The theoretical understanding on the direct CPV in charm decays then became confusing, ranging from $10^{-4}$ to $10^{-2}$ \cite{Grossman:2006jg,Buccella:1994nf,Grinstein,Bigi:2011re,Artuso:2008vf,FAT,FAT2,Cheng:2012wr,Brod:2011re,Pirtskhalava:2011va,
Brod:2012ud,Hiller:2012xm,Franco:2012ck,Feldmann:2012js,Khodjamirian:2017zdu,Muller:2015rna}. Among them, we predicted $\Delta A_{CP}\approx -1\times 10^{-3}$ using the factorization-assisted topological-amplitude (FAT) approach in the SM in 2012 \cite{FAT}, which is much smaller than the LHCb data in 2011 \cite{LHCb2011}. Considering the uncertainties of some theoretical inputs, such as $|V_{ub}V_{cb}^*|$, the $\gamma$ angle, the light quark masses and the masses and widths of scalar mesons, the prediction is \cite{FAT}
\begin{align}\label{eq:AcpFAT}
\Delta A_{CP}^{\rm SM}=(-0.57\sim-1.87)\times10^{-3}.
\end{align}
The first LHCb observation in Eq.(\ref{eq:AcpLHCb2019}) is very consistent with our prediction in Eq.(\ref{eq:AcpFAT}).

The dynamics of charm weak decays is difficult to calculate since the charm mass scale of $\mathcal{O}(1\text{GeV})$ is not high enough for the heavy quark expansion. To handle the significant non-perturbative contribution, the conventional approach to the analysis of charm decays is based on the topological-amplitude parametrization \cite{Chau:1982da,Chau:1986du}. For the tree amplitudes, benefitted from the abundant data of branching fractions, the non-perturbative contribution can then be extracted. However, the knowledge on the penguin amplitudes, to which branching ratios are not sensitive, is poor, making reliable predictions for CPV extremely challenging. The advantage of the FAT approach, compared to the conventional one, is that it serves as a framework, in which the flavor $SU(3)$ symmetry breaking effects can be included easily \cite{FAT,FAT2,Muller:2015lua,Paul}. It has been known that the $SU(3)$ symmetry breaking effects are crucial for explaining the dramatic difference between the $D^0\to K^+K^-$ and $D^0\to \pi^+\pi^-$ branching fractions, and the non-vanishing $D^0\to K^0\overline K^0$ branching fraction. Moreover, the FAT approach provides a prescription, through which the penguin amplitudes can be related to the tree ones as much as possible, such that predictions for CPV are possible. The details of the FAT approach can be found in \cite{FAT,FAT2}, and will not be repeated here. The FAT approach has been intensively applied to the studies of $D^0-\overline D^0$ mixing \cite{Jiang:2017zwr}, $K_S^0-K_L^0$ asymmetries \cite{Wang:2017ksn} and CPV \cite{Yu:2017oky} in charm decays into neutral kaons, and $B$ meson decays and their CPV \cite{Zhou:2015jba,Zhou:2016jkv,Wang:2017hxe}.

We intend to make the following remarks:
\begin{itemize}
\item The order of magnitude of charm CPV, $A_{CP}=\mathcal{O}(10^{-3})$, generally excludes the possibility of charm CPV at the percent level. Some other modes with CPV of order $10^{-3}$ are also expected.

\item The consistency of the experimental data in Eq.(\ref{eq:AcpLHCb2019}) with the theoretical prediction in Eq.(\ref{eq:AcpFAT}) indicates that the FAT approach is reasonable and reliable for estimating the penguin amplitudes. 
It is then likely to constrain new physics contributions to charm decays in the FAT framework.   

\item The precision of measurements on charm CPV is approaching $\sigma_{A_{CP}}=\mathcal{O}(10^{-4})$, at LHCb, which is really an amazing level of precision. The observation of charm CPV in other decay modes is promising. The FAT approach could help us to identify other golden channels to search for CPV in the charm sector.
\end{itemize}

Recent studies are performed in \cite{Xing:2019uzz}. In this paper, we will elaborate the implications of the observed charm CPV. In Sec.II, we explain how a CPV of order $10^{-3}$ can be well understood in the SM. In Sec. III, an experimental proposal is made to measure the $D^+\to K^+K^-\pi^+$ decay as the next potential channel for observing charm CPV. The summary is given in Sec. IV.

\section{Understanding of CPV in the SM}

The observation of charm CPV by LHCb with $\Delta A_{CP}$ at the order of $10^{-3}$ may be questioned to be understandable in the SM, or a signal of new physics. The naive expectation in Eq.(\ref{eq:naiveAcp}) is $\mathcal{O}(10^{-4})$. In this section, we will show how the CPV of order $10^{-3}$ can be well accommodated in the SM, viewing the consistency between the data in Eq.(\ref{eq:AcpLHCb2019}) and the SM prediction in Eq.(\ref{eq:AcpFAT}) in the FAT approach.

We start from the effective Hamiltonian 
\begin{align}
 \mathcal H_{\rm eff}={G_F\over \sqrt 2}
 \left[\sum_{q=d,s}V_{cq}^*V_{uq}(C_1(\mu)O_1^q(\mu)+C_2(\mu)O_2^q(\mu))
 -V_{cb}^*V_{ub}\left(\sum_{i=3}^6C_i(\mu)O_i(\mu)+C_{8g}(\mu)O_{8g}(\mu)\right)\right],
 \end{align}
 where $G_F$ is the Fermi coupling constant,
$V_{i}$'s denote the Cabibbo-Kobayashi-Maskawa
(CKM) matrix elements, and $C_{i}$'s are the Wilson coefficients.
The current-current operators are
\begin{eqnarray}
& &O_1=(\bar{u}_{\alpha}q_{2\beta})_{V-A}
(\bar{q}_{1\beta}c_{\alpha})_{V-A},
~~~~O_2=(\bar{u}_{\alpha}q_{2\alpha})_{V-A}
(\bar{q}_{1\beta}c_{\beta})_{V-A},
\end{eqnarray}
with the QCD penguin operators
 \begin{eqnarray}
 O_3&=&\sum_{q'=u,d,s}(\bar u_\alpha c_\alpha)_{V-A}(\bar q'_\beta
 q'_\beta)_{V-A},~~~
 O_4=\sum_{q'=u,d,s}(\bar u_\alpha c_\beta)_{V-A}(\bar q'_\beta q'_\alpha)_{V-A},
 \nonumber\\
 O_5&=&\sum_{q'=u,d,s}(\bar u_\alpha c_\alpha)_{V-A}(\bar q'_\beta
 q'_\beta)_{V+A},~~~
 O_6=\sum_{q'=u,d,s}(\bar u_\alpha c_\beta)_{V-A}(\bar q'_\beta
 q'_\alpha)_{V+A},
 \end{eqnarray}
and the chromomagnetic penguin operator,
\begin{eqnarray}
O_{8g}=\frac{g}{8\pi^2}m_c{\bar
u}\sigma_{\mu\nu}(1+\gamma_5)T^aG^{a\mu\nu}c,
\end{eqnarray}
$T^a$ being the color matrix. The explicit expressions and values of the Wilson coefficients $C_{1\text{-}6,8g}(\mu)$ are referred to \cite{FAT}.

The amplitudes of the  $D^0\to K^+K^-$ and $\pi^+\pi^-$ decays are written as
\begin{align}
\mathcal{A}(D^0\to K^+K^-)&=\lambda_s \mathcal{T}^{KK} +\lambda_b \mathcal{P}^{KK},\\
\mathcal{A}(D^0\to \pi^+\pi^-)&=\lambda_d \mathcal{T}^{\pi\pi} +\lambda_b \mathcal{P}^{\pi\pi},
\end{align}
where $\lambda_i=V_{ui}V_{ci}^*$, $i=d,s,b$, and $\mathcal{T}$ and $\mathcal{P}$ are the tree and penguin amplitudes, respectively. 
The $d$ and $s$ quark loops can be absorbed into the above terms using the CKM unitarity relation. 
The difference of CP asymmetries between the above two modes is
\begin{align}
\Delta A_{CP}= -2 r\sin\gamma \left({|\mathcal{P}^{KK}|\over |\mathcal{T}^{KK}|}\sin\delta^{KK}+{|\mathcal{P}^{\pi\pi}|\over |\mathcal{T}^{\pi\pi}|}\sin\delta^{\pi\pi}\right),
\end{align}
where $r=|\lambda_b/\lambda_{d,s}|$, and $\delta$'s are relative strong phases between the tree and penguin amplitudes. The process-independent factor is $2r\sin\gamma=1.5\times10^{-3}$ \cite{PDG}. The measured CPV in Eq.(\ref{eq:AcpLHCb2019}) implies
\begin{align}
\left({|\mathcal{P}^{KK}|\over |\mathcal{T}^{KK}|}\sin\delta^{KK}+{|\mathcal{P}^{\pi\pi}|\over |\mathcal{T}^{\pi\pi}|}\sin\delta^{\pi\pi}\right)\approx1.
\end{align}
Under the flavor $SU(3)$ symmetry, all the above quantities for $KK$ and $\pi\pi$ should be equal, such that ${|\mathcal{P}|\over |\mathcal{T}|}\sin\delta\sim1/2$, or ${\rm Im}[\mathcal{P}/\mathcal{T}]\sim1/2$.
The FAT approach led to \cite{FAT}
\begin{align}\label{eq:PoverT}
{\mathcal{P}^{\pi\pi}\over\mathcal{T}^{\pi\pi}}=0.66e^{i134^\circ},~~~\text{and}~~~{\mathcal{P}^{KK}\over\mathcal{T}^{KK}}=0.45e^{i131^\circ},
\end{align}
namely, $\Delta A_{CP}=-1\times10^{-3}$ which is very close to the data in Eq.(\ref{eq:AcpLHCb2019}). It implies that the estimate of the penguin amplitudes in the FAT approach is reliable. 


The dominant penguin contributions come from the QCD-penguin amplitude $P$, and the penguin-exchange amplitude $PE$ \footnote{The symbol of  $P_C$ in \cite{FAT} denotes the amplitude $P$actually.}, as shown in Fig.1. 
\begin{figure}[!phtb]
\includegraphics[scale=0.3]{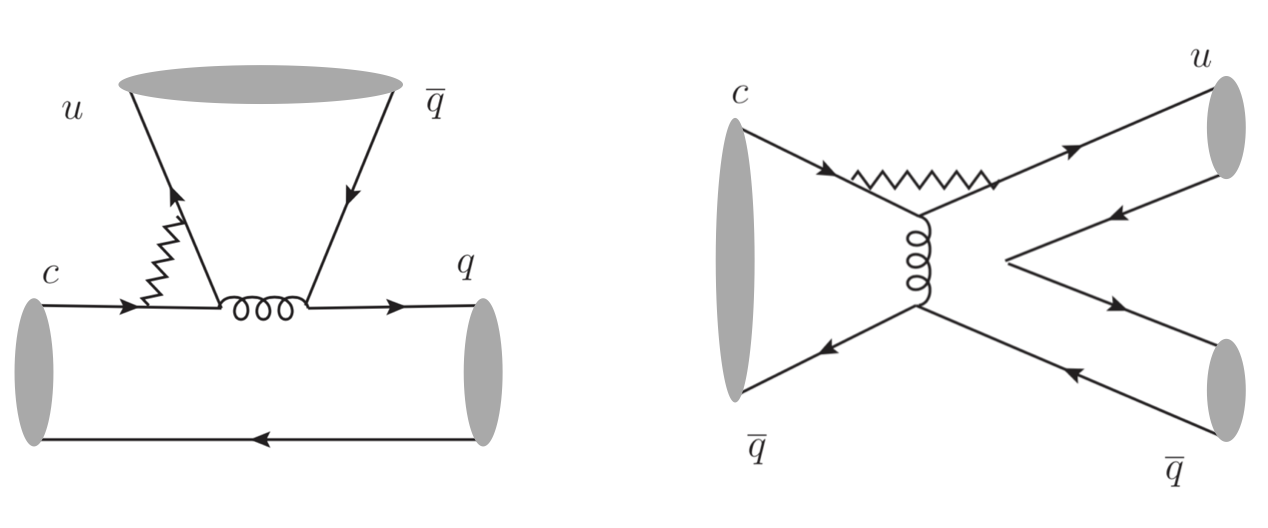}
\caption{Dominant penguin diagrams in charm decays.}
\end{figure}
%
The $d$ and $s$ quark loops and the chromomagnetic-penguin contributions are absorbed into the Wilson coefficients of \cite{FAT,FAT2}
\begin{align}
C_{3,5}(\mu)&\to C_{3,5}-\frac{\alpha_s(\mu)}{8\pi N_c}
\sum_{q=d,s}{\lambda_q\over\lambda_b}C^q(\mu,\langle l^2\rangle)
+\frac{1}{N_c}\frac{\alpha_s(\mu)}{4\pi}\frac{m_c^2}{\langle l^2\rangle}
[C_{8g}(\mu)+C_5(\mu)],\\
C_{4,6}(\mu)&\to C_{4,6}+\frac{\alpha_s(\mu)}{8\pi }
\sum_{q=d,s}{\lambda_q\over\lambda_b}C^q(\mu,\langle l^2\rangle)
-\frac{\alpha_s(\mu)}{4\pi}\frac{m_c^2}{\langle l^2\rangle}[C_{8g}(\mu)+C_5(\mu)],
\end{align}
with $C^q(\mu,\langle l^2\rangle)=\left[-\frac{2}{3}-4\int_0^1dxx(1-x)\ln\frac{m_q^2-x(1-x)
\langle l^2\rangle}{\mu^2}\right]C_2(\mu)$, and $\langle l^2\rangle\approx m_D^2/4$ in an assumption that each spectator of a light meson carries half of the meson momentum. At $\mu=1$GeV, we have $a_4=-0.036-i0.098$, $a_6=-0.031-i0.098$, i.e., $|a_4|\sim|a_6|\sim0.1$. The ratio of the QCD-penguin amplitude $P$ over the color-favored tree-emission $T$ without the CKM matrix elements is given by 
\begin{align}\label{eq:PoT}
 {P\over T}= {a_4+a_6 r_\chi\over a_1}=0.36 e^{-i 108^\circ}.
\end{align}
with the chiral factor $r_\chi=2m_0^2/m_c=2.8$, where $m_0^\pi=m_\pi^2/(m_u+m_d)$ and $m_0^K=m_K^2/(m_s+m_u)$.
Even if considering only the factorizable $T$ and $P$ contributions, we get $\Delta A_{CP}=-1\times10^{-3}$. Thus a charm CPV of the order $10^{-3}$ can be well understood. The penguin-exchange amplitude further enhances the penguin contributions, as seen by comparing Eq.(\ref{eq:PoT}) and Eq.(\ref{eq:PoverT}).

\section{Experimental Proposal}
After the observation of CPV via the $D^0\to K^+K^-$ and $\pi^+\pi^-$ channels, an important issue is which process is the next one to observe the charm CPV. It will be a good hint that the precision of measurement on $\Delta A_{CP}$ is $\mathcal{O}(10^{-4})$. The current largest data set is from LHCb, which will not be changed in the near future even though Belle II is starting operation. Hence, we consider only the processes available at LHCb, especially with charged final states. We find that the $D^+\to K^+K^-\pi^+$ mode is of high interest with the branching fraction \cite{PDG}
\begin{align}
\mathcal{B}(D^+\to K^+K^-\pi^+)&=(9.51\pm0.34)\times 10^{-3},
\end{align}
which is the largest one comparing the singly Cabibbo-suppressed (SCS) processes with all charged particles in the final states: it is twice larger than $D^0\to K^+K^-$, and six times larger than $D^0\to \pi^+\pi^-$. It is then expected that the precision of measurement on CPV of the above mode could reach the order of $10^{-4}$, similarly to or even better than $\Delta A_{CP}$. The search for CPV in $D^+\to K^+K^-\pi^+$ has been performed by LHCb and BaBar \cite{Aaij:2011cw,Lees:2012nn}, but with no signal of CPV due to the limited data samples. It deserves study with the full data of RUN I+II at LHCb. 

It has been shown that $D^+\to K^+K^-\pi^+$ is dominated by the quasi-two-body decays via $\phi\pi^+$, $\overline K^{*0}K^+$ and $K^+\overline K_0^*(1430)^0$, up to around $75\%$ of the total rate. The other resonant or non-resonant contribution is less than $10\%$ of the fit fractions, which can be safely neglected. The CPV of $D^+\to \phi\pi^+$ and  $\overline K^{*0}K^+$ have been predicted in the FAT approach \cite{FAT2} with values of $-1\times10^{-7}$ and $2\times10^{-4}$, respectively. In this work, we will examine the CPV in $D^+\to K^+\overline K_0^*(1430)^0$.

The tree amplitude of $D^+\to K^+\overline K_0^*(1430)^0$ contains the color-favored tree-emission diagram $T$ and the $W$-annihilation diagram $A$. In the $PP$ and $PV$ modes, the $A$ diagrams are always much smaller than the $T$ diagrams \cite{FAT,FAT2,Cheng:2010ry,Cheng:2016ejf}. We thus neglect the diagram $A$ in this analysis which may not affect the prediction for CPV very much. The diagram $T$ can be calculated in the factorization approach,
\begin{align}
T={G_F\over\sqrt2}V_{cs}^*V_{us}a_1(\mu)f_K(m_D^2-m_{K_0^*}^2)F_0^{D\to K_0^*}(m_K^2).
\end{align}

The penguin contribution to $D^+\to K^+\overline K_0^*(1430)^0$ is estimated below. To catch the dominant contributions, we consider only the QCD-penguin diagram $P$ and the penguin-exchange diagram $PE$, as discussed in the previous section. Both are dominated by factorizable contributions. The amplitude $P$ with transition to a scalar meson and emission of a pseudoscalar meson, Fig.1(left), is expressed as
\begin{align}
P&=-{G_F\over\sqrt2}V_{cb}^*V_{ub}\big[a_4(\mu)\langle K|(\bar u s)_{V-A}|0\rangle\langle K_0^*|(\bar s c)_{V-A}|D\rangle -2a_6(\mu)\langle K|(\bar u s)_{S+P}|0\rangle\langle K_0^*|(\bar s c)_{S-P}|D\rangle\big]
\nonumber\\
&={G_F\over\sqrt2}V_{cb}^*V_{ub}\big[a_4(\mu)+a_6(\mu)r_\chi\big]f_K(m_D^2-m_{K_0^*}^2)F_0^{D\to K_0^*}(m_K^2),
\end{align}
with the Wilson coefficients $a_4=C_4+C_3/N_c$ and $a_6=C_6+C_5/N_c$, and the chiral factor $r_\chi=2m_K^2/(m_c(m_s+m_u))$. The $D\to K_0^*$ transition form factor has been derived in \cite{Cheng:2003sm,Cheng:2010vk}.

The penguin-exchange diagram in Fig.1(right) with the $(S+P)(S-P)$ operator, dominated by the factorizable contribution, is evaluated using the pole model in \cite{FAT,FAT2}:
%
\begin{align}
PE&=-{G_F\over\sqrt2}V_{cb}^*V_{ub}a_6(\mu)(-2)\langle K^+\overline K_0^*(1430)^0 |P^* \rangle{i\over m_D^2-m_{P^*}^2}\langle P^*| (\bar u d)_{S+P}|0\rangle\langle 0|(\bar d c)_{S-P}|D\rangle
\nonumber\\
&= {G_F\over\sqrt2}V_{cb}^*V_{ub}a_6(\mu)r_\chi g_S f_{P^*} f_D m_D^2{1\over m_D^2-m_{P^*}^2},
\end{align}
where the strong coupling $g_S=3.8$ GeV is extracted from the $K_0^*(1430)\to K\pi$ data \cite{Fusheng:2011tw}. The pole here is a pseudoscalar meson, thus chosen as a pion. 

In the end, the CPV in the $D^+\to K^+\overline K_0^*(1430)^0$ decay is predicted as
\begin{align}
A_{CP}(D^+\to K^+\overline K_0^*(1430)^0)=-0.88\times10^{-3}.
\end{align}
Such a value of order $10^{-3}$ can be observed by LHCb with a precision of measurement at the order of $10^{-4}$. 

\section{Summary}
In summary, the observation of charm CPV is a milestone of high energy physics. The data, consistent with our prediction in the FAT approach in \cite{FAT}, indicate that the penguin dynamics in charm decays can be well estimated in the SM. The precision of measurements on charm CPV is approaching $\sigma_{A_{CP}}=\mathcal{O}(10^{-4})$, at LHCb, so the observation of charm CPV in other modes is promising. In this short paper we have obtained $A_{CP}(D^+\to K^+\overline K_0^*(1430)^0)=-0.88\times10^{-3}$, and proposed that the $D^+\to K^+K^-\pi^+$ decay might be the next potential channel to observe charm CPV. 


\section*{Acknowledgements}

This work was supported in part by the Ministry of Science and Technology
of R.O.C. under Grant No. MOST-107-2119-M-001-035-MY3, and
by the National Natural Science Foundation of China under the Grant No.11521505, 11575005, 11621131001 and U1732101.

\end{document}